\documentstyle[twoside]{article}

\catcode`\@=11
\long\def\@makefntext#1{
\protect\noindent \hbox to 3.2pt {\hskip-.9pt
$^{{\eightrm\@thefnmark}}$\hfil}#1\hfill}               

\def\thefootnote{\fnsymbol{footnote}}
\def\@makefnmark{\hbox to 0pt{$^{\@thefnmark}$\hss}}    

\def\ps@myheadings{\let\@mkboth\@gobbletwo
\def\@oddhead{\hbox{}
\rightmark\hfil\eightrm\thepage}
\def\@oddfoot{}\def\@evenhead{\eightrm\thepage\hfil
\leftmark\hbox{}}\def\@evenfoot{}
\def\sectionmark##1{}\def\subsectionmark##1{}}

\oddsidemargin=\evensidemargin
\renewcommand{\thefootnote}{\fnsymbol{footnote}}

\newcounter{sectionc}\newcounter{subsectionc}\newcounter{subsubsectionc}
\renewcommand{\section}[1] {\vspace{12pt}\addtocounter{sectionc}{1}
\setcounter{subsectionc}{0}\setcounter{subsubsectionc}{0}\noindent
        {\tenbf\thesectionc. #1}\par\vspace{5pt}}
\renewcommand{\subsection}[1] {\vspace{12pt}\addtocounter{subsectionc}{1}
        \setcounter{subsubsectionc}{0}\noindent
        {\bf\thesectionc.\thesubsectionc. {\kern1pt \bfit #1}}\par\vspace{5pt}}
\renewcommand{\subsubsection}[1] {\vspace{12pt}\addtocounter{subsubsectionc}{1}
        \noindent{\tenrm\thesectionc.\thesubsectionc.\thesubsubsectionc.
        {\kern1pt \tenit #1}}\par\vspace{5pt}}
\newcommand{\nonumsection}[1] {\vspace{12pt}\noindent{\tenbf #1}
        \par\vspace{5pt}}

\newcounter{appendixc}
\newcounter{subappendixc}[appendixc]
\newcounter{subsubappendixc}[subappendixc]
\renewcommand{\thesubappendixc}{\Alph{appendixc}.\arabic{subappendixc}}
\renewcommand{\thesubsubappendixc}
        {\Alph{appendixc}.\arabic{subappendixc}.\arabic{subsubappendixc}}

\renewcommand{\appendix}[1] {\vspace{12pt}
        \refstepcounter{appendixc}
        \setcounter{figure}{0}
        \setcounter{table}{0}
        \setcounter{lemma}{0}
        \setcounter{theorem}{0}
        \setcounter{corollary}{0}
        \setcounter{definition}{0}
        \setcounter{equation}{0}
        \renewcommand{\thefigure}{\Alph{appendixc}.\arabic{figure}}
        \renewcommand{\thetable}{\Alph{appendixc}.\arabic{table}}
        \renewcommand{\theappendixc}{\Alph{appendixc}}
        \renewcommand{\thelemma}{\Alph{appendixc}.\arabic{lemma}}
        \renewcommand{\thetheorem}{\Alph{appendixc}.\arabic{theorem}}
        \renewcommand{\thedefinition}{\Alph{appendixc}.\arabic{definition}}
        \renewcommand{\thecorollary}{\Alph{appendixc}.\arabic{corollary}}
        \renewcommand{\theequation}{\Alph{appendixc}.\arabic{equation}}
        \noindent{\tenbf Appendix \theappendixc #1}\par\vspace{5pt}}
\newcommand{\subappendix}[1] {\vspace{12pt}
        \refstepcounter{subappendixc}
        \noindent{\bf Appendix \thesubappendixc. {\kern1pt \bfit #1}}
        \par\vspace{5pt}}
\newcommand{\subsubappendix}[1] {\vspace{12pt}
        \refstepcounter{subsubappendixc}
        \noindent{\rm Appendix \thesubsubappendixc. {\kern1pt \tenit #1}}
        \par\vspace{5pt}}

\newcommand{\textlineskip}{\baselineskip=13pt}
\newcommand{\smalllineskip}{\baselineskip=10pt}

\def\eightcirc{
\begin{picture}(0,0)
\put(4.4,1.8){\circle{6.5}}
\end{picture}}
\def\eightcopyright{\eightcirc\kern2.7pt\hbox{\eightrm c}}

\newcommand{\copyrightheading}[1]
        {\vspace*{-2.5cm}\smalllineskip{\flushleft
        {\footnotesize International Journal of Modern Physics B, #1}\\
        {\footnotesize $\eightcopyright$\, World Scientific Publishing
         Company}\\
         }}

\def\abstracts#1#2#3{{
        \centering{\begin{minipage}{4.5in}\baselineskip=10pt\footnotesize
        \parindent=0pt #1\par
        \parindent=15pt #2\par
        \parindent=15pt #3
        \end{minipage}}\par}}

\renewenvironment{thebibliography}[1]                   
        {\frenchspacing
         \ninerm\baselineskip=11pt
         \begin{list}{\arabic{enumi}.}
        {\usecounter{enumi}\setlength{\parsep}{0pt}
         \setlength{\leftmargin 12.7pt}{\rightmargin 0pt} 
         \setlength{\itemsep}{0pt} \settowidth
        {\labelwidth}{#1.}\sloppy}}{\end{list}}

\newcounter{itemlistc}
\newcounter{romanlistc}
\newcounter{alphlistc}
\newcounter{arabiclistc}

\newcommand{\fcaption}[1]{
        \refstepcounter{figure}
        \setbox\@tempboxa = \hbox{\footnotesize Fig.~\thefigure. #1}
        \ifdim \wd\@tempboxa > 5in
           {\begin{center}
        \parbox{5in}{\footnotesize\smalllineskip Fig.~\thefigure. #1}
            \end{center}}
        \else
             {\begin{center}
             {\footnotesize Fig.~\thefigure. #1}
              \end{center}}
        \fi}

\newcommand{\tcaption}[1]{
        \refstepcounter{table}
        \setbox\@tempboxa = \hbox{\footnotesize Table~\thetable. #1}
        \ifdim \wd\@tempboxa > 5in
           {\begin{center}
        \parbox{5in}{\footnotesize\smalllineskip Table~\thetable. #1}
            \end{center}}
        \else
             {\begin{center}
             {\footnotesize Table~\thetable. #1}
              \end{center}}
        \fi}

\def\@citex[#1]#2{\if@filesw\immediate\write\@auxout
        {\string\citation{#2}}\fi
\def\@citea{}\@cite{\@for\@citeb:=#2\do
        {\@citea\def\@citea{,}\@ifundefined
        {b@\@citeb}{{\bf ?}\@warning
        {Citation `\@citeb' on page \thepage \space undefined}}
        {\csname b@\@citeb\endcsname}}}{#1}}

\newif\if@cghi
\def\cite{\@cghitrue\@ifnextchar [{\@tempswatrue
        \@citex}{\@tempswafalse\@citex[]}}
\def\citelow{\@cghifalse\@ifnextchar [{\@tempswatrue
        \@citex}{\@tempswafalse\@citex[]}}
\def\@cite#1#2{{$\null^{#1}$\if@tempswa\typeout
        {IJCGA warning: optional citation argument
        ignored: `#2'} \fi}}

\def\pmb#1{\setbox0=\hbox{#1}
        \kern-.025em\copy0\kern-\wd0
        \kern.05em\copy0\kern-\wd0
        \kern-.025em\raise.0433em\box0}

\def\fnt#1#2{\footnotetext{\kern-.3em
        {$^{\mbox{\scriptsize #1}}$}{#2}}}

\def\fpage#1{\begingroup
\voffset=.3in
\thispagestyle{empty}\begin{table}[b]\centerline{\footnotesize #1}
        \end{table}\endgroup}

\def\runninghead#1#2{\pagestyle{myheadings}
\markboth{{\protect\footnotesize\it{\quad #1}}\hfill}
{\hfill{\protect\footnotesize\it{#2\quad}}}}
\font\tenrm=cmr10
\font\tenit=cmti10
\font\tenbf=cmbx10
\font\bfit=cmbxti10 at 10pt
\font\ninerm=cmr9

\font\eightrm=cmr8

\def\qed{\hbox{${\vcenter{\vbox{                        
   \hrule height 0.4pt\hbox{\vrule width 0.4pt height 6pt
   \kern5pt\vrule width 0.4pt}\hrule height 0.4pt}}}$}}

\renewcommand{\thefootnote}{\fnsymbol{footnote}}        

\def\bsc{{\sc a\kern-6.4pt\sc a\kern-6.4pt\sc a}}       
\def\bflatex{\bf L\kern-.30em\raise.3ex\hbox{\bsc}\kern-.14em
T\kern-.1667em\lower.7ex\hbox{E}\kern-.125em X}

\def\J{\mbox{Jac}}
\def\tp{2 \pi}
\def\nf{{1\over N!}}

\begin{document}

\runninghead{Gibbs' States for  Moser-Calogero Potentials}
{Gibbs' States for Moser-Calogero Potentials}

\normalsize\textlineskip
\thispagestyle{empty}
\setcounter{page}{1}

\copyrightheading{}                     

\vspace*{0.88truein}

\fpage{1}
\centerline{\bf GIBBS' STATES FOR  MOSER-CALOGERO POTENTIALS}
\vspace*{0.37truein}
\centerline{\footnotesize K.L. Vaninsky\footnote{
Permanent address: Department of Mathematics, Kansas State University,  
Manhattan, KS 66506, USA. Supported by NSF grant DMS-9501002.}}
\vspace*{0.015truein}
\centerline{\footnotesize\it School of Mathematics, Institute for Advance 
Study, }
\baselineskip=10pt
\centerline{\footnotesize\it Princeton, NJ 08940, USA}

\vspace*{0.21truein}
\abstracts{
We present two independent approaches for  computing the thermodynamics 
for classical particles interacting via the  
Moser--Calogero potential. Combining the results we propose the form of
equation of state or, what is equivalent, the asymptotics  of the Jacobian
between  volume
elements corresponding two symplectic structures on the phase space.
}{}{}



\vspace*{1pt}\textlineskip      
\section{Introduction}    
\vspace*{-0.5pt}
\noindent
Non-equilibrium statistical mechanics studies properties of dynamical systems
describing big or infinite systems of interacting particles.  The
Hamiltonian
structure of the equations  allows one to define on the phase space
so-called Gibbsian measures, invariant under the dynamics. These measures
are
usually constructed from the basic Hamiltonian(=energy) and other classical
conserved quantities such as the number of particles, integrals of the
momentum and the angular momentum.  They constitute a finite parameter
family, so-called grand--canonical ensemble.
At present there are  four known 1D hierarchies with
similar structure of the ensemble in the generic case and the same number
(2) of  integrable exceptions, as it is explained below.
 
Statistical mechanics of 1D particles is well studied, \cite{1}.
In the generic case of
Lenard-Jones potential the Gibbs' states form 3 parameter family.
In 1975-76,  Moser and Calogero discovered two
repulsive potentials $ U(y)=x^{-2} ,\;\;  \sinh^{- 2} x $
which are integrable by the inverse spectral transform, \cite{2,3}.
Additional Gibbs' states  from additional integrals of the motion different
from the classical ones for
such potentials were constructed by  Chulaevskii \cite{4}.
In the paper of  Gurevich \cite{5}, the exceptional
character of Moser--Calogero potentials was proved.
 
Statistical mechanics of the 1D semi-linear wave equation with
restoring force was studied by McKean--Vaninsky \cite{6,7}.
The phase space of the dynamical system is a product of two function
spaces. In this phase space, a Gibbsian  measure was constructed and the
existence of flow was  proved, as was the invariance of the measure
under the flow.
 
Also, two integrable equations were considered: linear Klein-Gordon and
Sinh-Gordon. These equations have integrals of local  densities as
an additional conserved quantities of  motion. These integrals produce
infinitely many additional Gibbs' states.  Klein and Sinh-Gordon
can be considered as
analogs of  Moser-Calogero potentials.

In such integrable equations the conserved quantities
prevent any Gibbs' state from being ergodic in the periodic case. On the
entire line, these are
destroyed on the support of the measure. As it was proved by McKean
\cite{8}, for the $\sinh$-Gordon equation,  the Gibbs'
state constructed from the basic
Hamiltonian is ergodic.
It is believed that all other mutually singular Gibbs' states are
ergodic too.

Another field model where it is possible to carry out this program  is
the defocusing nonlinear Schr\"{o}dinger equation with the power
nonlinearity, \cite{9,10}. Again, there are
two integrable equations,  the linear and the cubic  cases,
with additional Gibbs' states.
 
Recently, \cite{11}, another hierarchy of so-called modified KdV equations
with similar properties  was discovered. Now all these one-dimensional
hierarchies fit nicely into the table.

\begin{center}   
\setlength{\unitlength}{0.85 mm}
\begin{picture}(180,75)(15,0)

\put(0,0) {\framebox(43,15){$\frac{1}{q^2}\;\;\;\;\;  \frac{1}{\sinh^2 q}$}}
\put(0,15){\framebox(43,15){$e^{-\beta H - \alpha N - \gamma P}$}}
\put(0,30){\framebox(43,10){$N$=$\sharp$ of particles}}
\put(0,40){\framebox(43,10){$P= \sum p$=momentum}}
\put(0,50){\framebox(43,10){$H$=energy}}
\put(0,60){\framebox(43,15){$H=\sum \frac{p^2}{2}+\sum U(\Delta q)$}}

\put(43,0) {\framebox(43,15){$q^2 \;\;\;\; \cosh q$}}
\put(43,15){\framebox(43,15){$e^{-\beta H -  \gamma P}$}}
\put(43,30){\framebox(43,10){}} 
\put(43,40){\framebox(43,10){$P= \int pq'$=momentum}}
\put(43,50){\framebox(43,10){$H$=energy}}
\put(43,60){\framebox(43,15){$H =  \int \frac{p^2 + q'^2}{2} + F (q)$}}

\put(86,0) {\framebox(45,15){$|\psi|^2 \;\;\;\; |\psi|^4$}}
\put(86,15){\framebox(45,15){$e^{-\beta H - \alpha N - \gamma P}$}}
\put(86,30){\framebox(45,10){$N = \int | \psi |^2$=$\sharp$ of particles}}
\put(86,40){\framebox(45,10){$P= \int  \psi' \bar{\psi}$=momentum}}
\put(86,50){\framebox(45,10){$H$=energy}}
\put(86,60){\framebox(45,15){$H = \int | \psi' |^2 + F ( | \psi | ^2 )$}}

\put(131,0) {\framebox(43,15){$q^2 \;\;\;\; q^4$}}
\put(131,15){\framebox(43,15){$e^{-\beta H -  \gamma P}$}}
\put(131,30){\framebox(43,10){}}
\put(131,40){\framebox(43,10){$P= \int q^2$=momentum}}
\put(131,50){\framebox(43,10){$H$=energy}}
\put(131,60){\framebox(43,15){$H = \int \frac{q'^2}{2} + F (q)$}}

\end{picture}
\end{center}
 
The first line of the table displays the  Hamiltonians of
these four hierarchies.
For particles, the potential $U(\cdot)$ is assumed to be Lenard-Jones type.
For PDE hierarchies $F(\cdot)$ is the potential of a restoring force;
it is an even function increasing at infinity. The next three lines are
generic classical conserved quantities. The NLS-hierarchy and m-KdV
hierarchy do not have any analog of $N = \sharp$ of particles. The next line
is the symbol for the Gibbs' measure; it is a three/two parameter family.
The last line presents integrable potentials where additional conserved
quantities produce an additional Gibbs' states.
 
All three integrable nonlinear equations are so-called AKNS systems,
related to some selfadjoint problem for the Dirac operator. These are  all
the nonlinear
equations related to such  spectral problem. 
 
Additional structure of  integrability allows one to obtain
detailed information about the Gibbs' states. In  section  2 we describe
the decomposition of the canonical measure for the cubic Schr\"{o}dinger 
equation previously obtained
in \cite{11,12}. In section 3, taking such decomposition
for granted for  Moser--Calogero potentials, we  compute
the thermodynamic limit of the 
specific free energy, the average energy per particle and
the equation of state. In the last section 4 we present
an alternative approach in the case of the rational
potential $2 x^{-2}$ and formulate the conjecture on the explicit form
of  the equation of state and the asymptotics of the Jacobian.

\textheight=7.8truein
\setcounter{footnote}{0}
\renewcommand{\thefootnote}{\alph{footnote}}

\section{Gibbs' States in Action-Angle Variables.}
\noindent
For the cubic Scr\"{o}dinger equation, the global action-angle variables
$ 0\leq I_k,\; 0\leq \phi_k < 2\pi,$ $k=...,-1,0,1,...,$ 
associated with the basic symplectic structure
$\omega$   were constructed by McKean--Vaninsky \cite{11}.
The next step  is to express the measure
$$
e^{-H} d\, {\rm vol}, \quad \quad \mbox{where} \quad 
H=\int_{0}^{1} |\psi'|^2 + 2|\psi|^4
$$
in such coordinates.
The main ingredient is the trace formula $H=\sum I'_k$, where the $I'$s are
periods of some Abelian differential on the  Riemann surface.
Presumably they are the actions, but relative to some higher symplectic
structure $\omega'$.  On a formal level,
$$
e^{-H} d\, {\rm vol}= e^{-\sum I'}\;\;  \prod dI\, d\phi=
e^{-\sum I'} \;\;{\rm Jac} \;\;\prod dI'\, d\phi,
$$
where ${\rm Jac}$ is the Jacobian between the variables $I$ and $I'$. Therefore,
$$
{\rm Jac}^{-1}\, e^{-H} d\, {\rm vol}= e^{-\sum I'} \;\;\prod dI'\, d\phi,
$$
{\it i.e.} $I'$s and $\psi$'s are independent, the $I'$s are exponential and
the $\phi$'s are uniform.
Such a decomposition  was  proved by McKean--Vaninsky \cite{12}.
It can be useful in computing different thermodynamical quantities, as
explained below.

\section{Thermodynamics of Moser-Calogero Potentials.}
\noindent
Consider   particles
on the line interacting with the potential $U= 2 x^{-2}$ (the rational case) or
$2 \sinh^{-2}x$ (the trigonometric case).
To construct  the Gibbs' state for an  infinite system of such
particles,  we  periodize the potential as in 
$U_{\omega_1}(\bullet)=\sum\limits_{k} U(\bullet +k \omega_{1})$.
In our case, $U_{\omega}$ is simply the Weierstrass $2 \wp$ with possibly
infinite second imaginary period $\omega_2$ (the rational case).
Now $N$ particles on the circle of the perimeter $\omega_1$  are governed
by the Hamiltonian
$$
H(q,p)=\sum\limits_{k=1}^{N}{p_k^2\over 2} + \sum\limits_{k \neq j;k,j=1}^{N}
2 \wp(q_k-q_j).
$$
The partition function is defined as
$$
Z(N,\omega_1,\beta)=\nf \int\limits_{M} e^{-\beta H(q,p)} d^N q d^N p,
$$
where the integral is taken over the phase space $M=\left([0,\omega_1)\times
R^1\right)^N$ and $\beta=T^{-1}$ is the reciprocal temperature.
In the thermodynamic limit
$N$ and $\omega_1$ tend to $\infty$ , but their ratio $\omega_1/N$ tends to some
constant: the specific volume $v > 0$. It is well  known\cite{1} that, in
the limit,   the specific free energy per particle
$$
\psi(v,\beta)=\lim N^{-1}\log Z(N,\omega_1,\beta)
$$
exists in the limiting ensemble.
The average energy per particle  and the  pressure
can be computed by simple differentiation of $\psi$
$$
<E>=-{\partial \psi(v,\beta)\over \partial \beta}, \quad\quad\quad
p=T \; {\partial \psi(v,\beta)\over \partial v}.
$$
Assuming the possibility of   decompositioning for Gibbs' states 
described in section 2, we compute  $\psi(v,\beta)$ by some version of
the method of  stationary phase.
 
Obvious rotational symmetry allows one to reduce the dimension of the phase
space. Indeed, according to the classical prescription,  Arnold \cite{13},
$$
I_1= {1\over \tp}\int p\; dq= {1\over \tp}\int \sum p_k \;
dq_k ={\omega_1\over
\tp}
\sum p_k= {\omega_1\over \tp} P,
$$
where $P$ is the total momentum. The corresponding angle $\varphi_1$ is
just the usual  on the circle.
 
The  change of variables $q_k\rightarrow q_k'$ and
$p_k\rightarrow p_k' + {P\over N}$ makes the total momentum $P'$  vanish.
The construction of the action-angle variables
$I_2,\varphi_2, \cdots, I_N,\varphi_N$  on the reduced phase space is based
on the  KP--1 equation, see \cite{14}. We do not need an explicit representation
for them now. 

To  compute the partition function, the domain of integration is 
similarly reduced:
\begin{eqnarray*}
Z(N,\omega_1,\beta)& = \nf \int\limits_{M} e^{-\beta H(q,p)}
\int\limits_{R^1} \delta(P(q,p)-P_0) dP_0 \, d^Nq \, d^Np\\
&= \int\limits_{R^1}\, dP_0 \nf \int\limits_{M} e^{-\beta H(q,p)}
\delta(P(q,p)-P_0)  d^Nq \, d^Np.
\end{eqnarray*}
The canonical transformation $q_k\rightarrow q_k'$ and
$p_k\rightarrow p_k' + {P_0\over N}$ produces 
$$H(q,p)=H(q',p')+ {P_0^2\over 2N}, \quad\quad \quad\quad
P=P'+ P_0
$$ and
\begin{eqnarray*}
Z&=&\int\limits_{R^1} dP_0 \nf \int\limits_{M}
e^{-\beta H(q',p')-\beta P_0^2/2N}\delta (P') d^Nq'\;d^Np'\\
&=&\sqrt{\tp N T} \nf \int\limits_{M}e^{-\beta H(q,p)}\delta(P)\; d^Nq \; d^N p.
\end{eqnarray*}
Now make a canonical transformation to the action-angle variables
$I_1,\varphi_1,\cdots, I_N,\varphi_N$. Then 
$$
Z= \sqrt{\tp N T} \nf \int\limits_{\mbox{range of}\; I_1,\cdots, I_N
\times [0,\tp)^N} e^{-\beta H(I_2,\cdots, I_N)}\delta(I_1
{\tp\over \omega_1})\; d I_1\, d \varphi_1 \prod\limits_{k=2}^{N}
d I_k\; d \varphi_k.
$$
Integrating out $I_1$ and all the angles we obtain
$$
Z =\sqrt{\tp N T} \;{\omega_1\over \tp}\; (\tp)^N \nf
\int\limits_{\mbox{range of}\; I_2,\cdots, I_N }
e^{-\beta H(I_2,\cdots, I_N)} \prod\limits_{k=2}^{N} d I_k.
$$
To proceed farther we have to make  few assumptions. The arguments in
favor of them are supplied at the end of this section.
 
{\bf Assumption 1.} The energy $H$ can be expressed as
$H=\sum_{k=2}^N I_k'$ ("trace formula"),
where $I_2',\cdots,I_N'$ are the actions relative to some higher symplectic
structure $\omega'$.
 
{\bf Assumption 2.} The range of the variables $I_2', \cdots, I_N'$ is
the rectangular domain $\{(I_2',\cdots,I_N'):\;\; I_{k\min}'\leq I_k',
\quad k=2,\cdots, N\}$, as  shown in  the picture
 
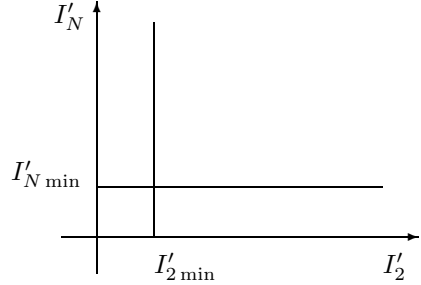
\begin{center}
\setlength{\unitlength}{.95mm}
\begin{picture}(80,40)(0,-5)
 
\put(20,0){\vector(1,0){50}}
\put(25,-5){\vector(0,1){38}}
\put(25,7){\line(1,0){40}}
\put(33,0){\line(0,1){30}}
 
\put(13,8){$I_{N\min}'$}
\put(19,30){$I_{N}'$}
 
\put(33,-5){$I_{2\min}'$}
\put(65,-5){$I_{2}'$}
 
\end{picture} 
\end{center}
Then $H_{\min}=\sum I'_{\min}$ and  in the thermodynamic limit,
$H_{\min}/N\rightarrow h(v)$, where $h(v)\equiv \sum_{k\neq 0} U(kv)$ is the
specific energy of one  particle in the ground state with the  specific 
volume $v$ of the limiting infinite system.
 
{\bf Assumption 3.} The Jacobian between the variables $I_2,\cdots,I_N$ and
 $I_2',\cdots,I_N'$ in the thermodynamic limit has the asymptotics
$$N^{-1} \log \J(I_2',\cdots,I_N', \omega_1,N)= \log N - s(v,\beta) + o(1)$$
in probability relative to the Gibbs' ensemble. 
The  nontrivial part of the asymptotics $s(v,\beta)$, we call 
{\it entropy}, by  analogy with  statistical mechanics,  Lanford \cite{15}.

Now we  employ the trace formula:
$$
Z(N,\omega_1,\beta)=\sqrt{\tp N T}\;  \omega_1 \; (\tp)^{N-1} \nf
\int\limits_{\mbox{range of}\; I_2',\cdots, I_N'}
e^{-\beta \sum I_k'} \; \J\,(I_2',\cdots,I_N')\;
\prod\limits_{k=2}^{N} d I_k'
$$
For the specific free energy in the thermodynamic limit, we obtain finally 
\begin{eqnarray*}
\psi(v,\beta)&=&\lim N^{-1}\log \;Z(N,\omega_1,\beta)\\
&=&\log  \tp + N^{-1}\log \nf \int\limits_{\mbox{range of}\; I_2',\cdots, I_N'}
e^{-\beta \sum I_k'} e^{N \log N -Ns(v,\beta)} \prod\limits_{k=2}^{N} 
d I_k' + o(1)\\
&=&\log \tp  - N^{-1}\log N! + N^{-1}\log  \int\limits_{\mbox{range of}\; 
I_2',\cdots, I_N'} 
e^{ -\beta \sum I_k'}  \prod\limits_{k=2}^{N} d I_k'  \\ 
&&+\log N -s(v,\beta)+ o(1) \\
&=&\log \tp  +1 + N^{-1}\log {e^{-\beta H_{\min}}\over \beta^{N-1}}
    - s(v,\beta)+ o(1)\\
&=&\log \tp +1 -\beta h(v) -\log \beta -s(v,\beta).
\end{eqnarray*}
The average energy is
$$
<E>=-{\partial \psi(v,\beta)\over \partial \beta}= T + h(v) +
 s_{\beta}'(v,\beta).
$$
The equation of state has the form
$$
p=T \; {\partial \psi(v,\beta)\over \partial v}= -T
s_{v}'(v,\beta) -h_{v}'(v).
$$
These formulas provide a bridge between two unsolved problems:
the asymptotics of the Jacobian determines  the thermodynamic
functions and {\it vice versa}. We will try to guess an explicit form of
$s(v,\beta)$ in the next section.

Now we explain why we expect our assumptions to be true.
 
{\bf Assumption 1.} The trace formula appears: in the linear problem, \cite{7},
as  the Parsevall identity and, in the cubic Schr\"{o}dinger case \cite {11},
as a result of contour integration on the Riemann surface. We expect such
formula here too, with $I'$s being  a periods of some Abelian differential.
 
{\bf Assumption 2.}  A general fact due to Atiyah-Guillemin-Sternberg
states that, for a   compact symplectic phase space with a torus
action,  the image of the momentum map is a convex polytope.
The shape of the momentum map here is suggested by the previous
work on  PDE and by the fact  that the energy in the ground
state is strictly positive.
 
{\bf Assumption 3.}  To explain why the $\J(I_2',\cdots,I_N', \omega_1,N)$
becomes a constant in thermodynamic limit, we have
to define all objects under consideration on one probability space. 
Consider a new
phase space $M$ for the  infinite-particle system on the entire line.
$M$ contains all possible positions and velocities of particles, with the
sole restriction  that  any finite region in the configuration  space contains
a finite number of particles (so-called locally finite).
 
Let $M_{\omega_1} \subset M$ comprise  all spatially $\omega_1$-periodic
configurations. On   $M_{\omega_1}$, let 
$\mu_{\omega_1}(\bullet)$ be a Gibbs' state with periodic boundary conditions.
Namely, consider projection of any configuration from
$M_{\omega_1}$ into a finite volume,  from $0$ to $\omega_1$. Any
configuration is determined by the projection. Define on the projections
a  Gibbs' state with periodic boundary condition. It
induces a  measure $\mu_{\omega_1}(\bullet)$ on    $M_{\omega_1}$ and on
the big space $M$, too.
A limiting Gibbs' state $\mu_{\infty}(\bullet)$ on $M$ is obtained from
$\mu_{\omega_1}(\bullet)$ by passing to the limit
$\omega_1\rightarrow \infty$.
 
On the big space $M$ acts the one-parameter group of spatial translations 
$\tau_{b}, b\in R^1$. Obviously, $\mu_{\omega_1}(\bullet)$ and
$M_{\omega_1} $ itself are invariant under the action of $\tau$. The limiting
measure $\mu_{\infty}(\bullet)$  is also invariant under such translations
and even ergodic.
 
For any such $M_{\omega_1}$ we can define a Jacobian
$\J(I_2',\cdots,I_N', \omega_1,N)$; it can be considered
on the whole $M$ as a function defined almost everywhere with
respect  to $\mu_{\omega_1}(\bullet)$.
The Jacobians are $\tau$--invariant functions. If they have a limit
defined almost everywhere with respect $\mu_{\infty}(\bullet)$, then
it is $\tau$-invariant and must be a constant due to the
ergodicity of this measure. The  exact form of the asymptotics is 
suggested by \cite{16}.

\section{Alternative Approach.  The Rational Case.}
\noindent
Gallavotti-Marchioro \cite{17} and also  Francoise \cite{18}, considered the  
system of N particles on the line with the Hamiltonian
$$
H(q,p)= \sum\limits_{k=1}^{N} {p_k^2\over 2} +{\lambda^2 q_k^2\over 2}
+ \sum\limits_{k \neq j;k,j=1}^{N} 2 (q_k-q_j)^{-2}.
$$
They obtained a closed expression for the partition function:
$$
Z(N,\lambda,\beta)=\nf \int\limits_{R^N\times R^N} e^{-\beta H} d^Nq\; d^Np=
\nf \left({2\pi\over \beta \lambda}\right)^N \exp(-\beta\lambda  N(N-1)).
$$

Now,  pass to the limit with $N\rightarrow \infty$ and assume
$\lambda \sim N^{-1}{c(v,\beta)},$ where $c(v,\beta)$ is some unknown
function. With this scaling  $H$ grow like $N$ in the
statistical ensemble. In the limit, we obtain an infinite--particle system
with    specific volume $v$ and temperature $\beta$.
For the specific free   energy we have
\begin{eqnarray*}
\psi(v,\beta)&=&  \phantom{-} N^{-1} \log \nf \left( {\tp N\over \beta
c}\right)^N e^{- \beta c (N-1)}+o(1)  \\
   &=& -N^{-1}\log N! + \log \tp + \log N -\log \beta -\log c -\beta c + o(1)\\ 
   &=& \phantom{-}\log \tp  +1 - \beta c -\log \beta -\log c .
\end{eqnarray*}
The expression for the specific free energy  contains the unknown
function $c(v,\beta)$, but the role of the two unknown functions 
$c(v,\beta)$ and $s(v,\beta)$ is different. Comparing these equations we   
conjecture $c(v,\beta)=h(v)$ and  $s(v,\beta)=\log h(v) $. In the rational
case $h(v)$ can be computed explicitly
$$
h(v)=\sum_{k\neq 0} U(kv) =\sum_{k\neq 0} {2\over k^2 v^2} =
{2\pi^2\over 3v^2}.
$$
Therefore
$$
p =  -T {\partial\over \partial v} \log h(v)-h_{v}'= {2T\over v} +
           {4\pi^2\over 3v^3}, \quad \quad \quad
<E> = T+h(v)= T + {2\pi^2\over 3v^2}.
$$
These equations are  similar to the  ideal gas for which $p=T/v$ and $<E>=T/2$.
In the trigonometric case ($U(x)=2\sinh^{-2}x$) we expect similar results.

\nonumsection{Acknowledgments}
\noindent
I would like to thank I.M. Krichever and  H.P. McKean for encouragement and 
numerous discussions. I am also grateful to Ya.G. Sinai and 
S.R.S. Varadhan  for useful remarks.

\nonumsection{References}

\end{document}